\documentclass[structabstract]{aa}  
\usepackage{graphicx}
\usepackage{txfonts}
\begin{document}
   \title{Galaxy interactions I: Major and minor mergers}

   \subtitle{}

   \author{Diego G. Lambas\inst{1,3}
   		  \and
          Sol Alonso\inst{1,2}
          \and
          Valeria Mesa\inst{1,2}
          \and
          Ana Laura O'Mill\inst{1,3}
          }

   \institute{Consejo Nacional de Investigaciones Cient\'{\i}ficas y
T\'ecnicas, Argentina\\
              \email{dgl@mail.oac.uncor.edu}
         \and
             ICATE, CONICET, CC 49, 5400 San Juan, Argentina\\
            \and
IATE, CONICET, OAC, Universidad Nacional de C\'ordoba, Laprida 854,
X5000BGR, C\'ordoba, Argentina
             }
             
   \date{Received xxx; accepted xxx}

  \abstract
   {}
   {We study galaxy pair samples selected from the Sloan Digital Sky Survey (SDSS-DR7) and we perform an analysis of minor and major mergers with the aim of investigating the dependence of galaxy properties on interactions. }
   {We build a galaxy pair catalog requiring $r_p < 25$ kpc $h^{-1}$ and 
$\Delta V < 350$ km $s^{-1}$ within redshift $z<0.1$.
By visual inspection of SDSS images we removed false identifications and we classify the interactions into three categories: pairs undergoing merging, $M$; pairs with evident tidal features, $T$; and non disturbed, $N$.
We also divide the pair sample into minor and major interactions according to the 
luminosity ratio of the galaxy members. We study star formation activity through colors, the 4000 $\AA$ break, and star formation rates.}
   { We find that $\sim 10\%$ of the pairs are classified as $M$. These systems show an excess of young 
stellar populations as inferred from the $D_n(4000)$ spectral index, colors, and star formation 
rates of the member galaxies, an effect which we argue, is directly related to the ongoing merging process. 
We find $\sim 30\%$ of the pairs exhibiting tidal features ($T$ pairs) with member galaxies
showing evidence of old stellar populations. This can be associated either to the disruptive effect of some tidal interactions, or  to the longer time-scale of morphological 
disturbance with respect to the bursts of the tidal induced star formation. \\ 
Regardless of the color distribution, we find a prominent blue peak in the strongest mergers, while pairs with tidal signs under a minor merger show a strong red peak. Therefore, our results show that galaxy interactions are important in driving the evolution of galaxy bimodality.\\
By adding stellar masses and star formation rates of the two members of the pairs, we explore the global 
efficiency of star formation of the pairs as a whole. We find that, at a given total stellar mass, major 
mergers are significantly more efficient (a factor $\approx$ 2) in forming new stars, with respect to 
both minor mergers or a control sample of non-interacting galaxies.

We conclude that the characteristics of the interactions and the ratio of luminosity galaxy pair members 
involved in a merger are important parameters in setting galaxy properties.
}
   {}


   \maketitle
%

\section{Introduction}

  Over the history of the universe, galaxy-galaxy interactions link the process of star formation 
with the growth of galaxies. According to hierarchical structure formation models, these interactions 
play a critical role in the formation and evolution of galaxies as discussed by Woods et al. 2007 and 
references there in. Simulations show that galaxies grow by accreting other galaxies, mostly minor 
companions. Although collision of comparable galaxies are expected to be the most damaging, encounters between galaxies and minor companions should be the most common type of interaction because of the greater fractional abundance of low luminosity galaxies.\\
Analysis of observational data have also shown that galaxy interactions are powerful mechanisms to trigger star formation (Yee \& Ellingson 1995; Kennicutt 1998). Barton, Geller \& Kenyon (2000) and Lambas et al. (2003) have carried out statistical analysis of star formation activity of galaxy pairs, finding that proximity in radial velocity 
and projected distance is correlated to an increase of the star formation activity.\\
The underling physics of star formation activity during galaxy-galaxy interactions have been explained 
by both theoretical (Martinet 1995), and numerical analysis (e.g. Toomre \& Toomre 1972; Barnes \& Hernquist 1992, 1996; Mihos \& Hernquist 1996). 
These studies showed that starbursts are fueled by gas inflows produced by the tidal torques generating during the encounters.
The efficiency of this mechanism depends on the particular internal characteristics of galaxies and their gas reservoir.
In pairs with similar luminosity galaxies (i.e. major interactions) there is an important redistribution of mass and a strong gravitational tidal torque causing gas angular momentum to be transferred outwards before the final merger.
In pairs formed by two galaxies with a large relative luminosity ratio (i.e. minor interactions) the tidal action from the less massive companions can induce a non-axisymmetric structure in the disk of the main galaxy 
(Hernquist \& Mihos 1995).  
The star formation activity in minor interactions depend on structural and orbital parameters. Close passage, prograde orbits between bulge-less galaxies are the most efficient at inducing gas inflows, and therefore trigger the star formation (Cox 2009).

On the observational side, Woods, Geller \& Barton (2006) analyzed a sample of 
136 pairs from the CfA2 Redshift Survey. The authors find that the relative luminosity of the companion 
galaxy is a determinant parameter of the star formation activity induced by the tidal effects of the interaction.
Also, Woods \& Geller (2007) found that, in minor pairs, the faint galaxy member shows evidence for tidally triggered star formation, whereas the primary galaxy is in general not strongly affected by the interaction. On the other hand, both galaxies undergoing a major interaction show enhanced star formation.
Michel-Dansac et al. (2008) studied the metallicity in a sample of galaxy pairs taken from SDSS-DR4, 
finding that, in minor interactions, the less massive galaxy member is systematically enriched, 
while  interacting with a comparable stellar mass companion shows a systematic metallicity decrement with
respect to galaxies in isolation. The authors argue that metal-rich starbursts triggered by a more massive component, and inflows of low metallicity gas induced by comparable or less massive companion galaxies are a natural explanation for these results.
By studying the spectral index, $D_n(4000)$, and star formation rates, $SFR$, Woods et al. (2010) showed the presence of bursts of star formation associated to major galaxy interactions, in particular, in very close pairs. Patton et al. (2010) studied optical colors finding an important fraction of red galaxies in pairs, a somewhat expected result since their galaxy pair sample reside preferentially in higher density environments than non-paired galaxies. These authors also found clear signs of interaction-induced star formation in the blue galaxies of close pairs.
Robaina et al. (2009) studied the dependence of $SFR$ on projected galaxy separation
using COMBO-17 data finding that only 10$\%$ of the star formation at intermediate
redshift is triggered directly by major mergers and interactions.
More recently, Darg et al. (2010) present a catalog of 39 multiple-mergers at $z<0.1$ from the merger catalog
of the Galaxy Zoo project where the member objects have properties typical of elliptical galaxies.

In this paper we focus on a statistical analysis of close galaxy pairs defined as those with relative projected separation  
$r_{p}< 25 \rm \,kpc \,h^{-1}$ and relative radial velocities $\Delta V < 350 \rm \,km \,s^{-1}$. According to these prescriptions, we have constructed a catalog of close pairs from the SDSS-DR7 and following 
Alonso et al. (2007) these pairs were classified according to the level of morphological disturbance associated to the interaction. 
In an attempt to explore the physical mechanisms that may affect the star formation activity, we have analyzed galaxy luminosities, spectral indicators of stellar populations, and colors. We use K-corrections of the publicly available code described in Blanton \& Roweis (2007) (\texttt{k-correct\_v4.2}) as a calibration for our k-corrected magnitudes.

This paper is structured as follows: Section 2 describes the procedure used to construct the catalog 
and explain the process of classification and depuration of pair galaxies. In Section 3 we study and characterize the effects of major and minor interactions and we discuss the dependence of star formation on colors and stellar population age. In Section 4, we summarize our main conclusions.


\section{Construction of a galaxy pair catalog from SDSS-DR7}

The Sloan Digital Sky Survey (SDSS, York et al. 2000) is the largest galaxy 
survey at the present.
It uses a 2.5m telescope (Gunn et al. 2006) to obtain photometric and 
spectroscopy data that will cover approximately one-quarter of 
the celestial sphere and collect spectra of more than one million objects.
The seven data release imaging (DR7, Abazajian et al. 2009) comprises 11663 
square degrees of sky imaged in five wave-bands ($u$, $g$, $r$, $i$ and $z$) containing
photometric parameters of 357 million objects. 
Within the survey area, DR7 includes spectroscopic data covering 9380 square 
degrees with 929555 spectra of galaxies.
DR7 therefore represents the final data 
set released with the original targeting and galaxy selection 
(Eisenstein et al. 2001, Strauss et al. 2002).
The main galaxy sample is essentially a magnitude limited spectroscopic sample 
(Petrosian magnitude) \textit{$r_{lim}$}$ < 17.77$, most of galaxies span a
redshift range $0 < z < 0.25$ with a median redshift of 0.1 (Strauss et al. 2002).
We considered a shorter redshift range, $z<0.1$, in order to avoid 
strong incompleteness at larger distances (Alonso et al. 2006). 

We build a Galaxy Pair Catalog (GPC) from the SDSS-DR7, following our previous works 
(Alonso et al. 2007),
requiring members to have relative projected separations, $r_{p}< 25 \rm \,kpc \,h^{-1}$
and relative radial velocities, $\Delta V< 350 \rm \,km \,s^{-1}$ within redshifts $z<0.1$. 
The number of pairs satisfying these criteria is 5579.
We exclude AGNs for 
our sample, which could affect our interpretation 
of the results due to contributions from their emission line spectral features. We have 
also removed false identifications (mostly parts of the same galaxy) and objects with large magnitude uncertainties. 
With these restrictions, our final pair catalog in the SDSS-DR7 comprises 1959 reliable close galaxy pairs with $z<0.1$. In Table 1 we summarize this results.
As discussed in Alonso et al. (2006), the effects of incompleteness or 
aperture (e.g see also Balogh et al.2004) do not
introduce important bias in the galaxy pair catalogs.

\begin{table}
\center
\caption{Percentages of different galaxy pair types}
\begin{tabular}{|c c c | }
\hline
Pair types &  Number of pairs  &  Percentages  \\
\hline
\hline
Pairs &  5579   &   100$\%$    \\
\hline
AGN-AGN pairs  &  370   &    6.63$\%$    \\
AGN-Galaxy pairs  &  1324   &    23.73$\%$    \\
False pair identification  & 1788  & 32.05$\%$    \\
Uncertainty magnitude pairs & 138  & 2.47$\%$    \\
\hline
Real Galaxy-Galaxy pairs  & 1959    &  35.11$\%$    \\          
\hline
\end{tabular}
{\small}
\end{table}

\subsection{Classification of galaxy pairs}

\begin{figure}
\includegraphics[width=95mm,height=30mm ]{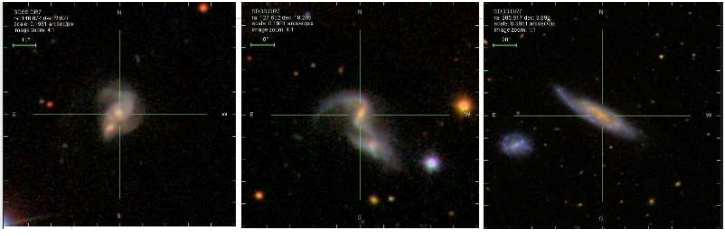}
\caption{Examples of galaxy pair images with different classification: 
$M$ (left panel), z=0.060; $T$ (medium panel), z=0.027 and $N$, z=0.023(right panel); the scale, size in arcsec and the N-E direction can be seen in Figure.}
\label{ej}
\end{figure}

We classified all galaxies in the pair catalog taking into account the 
eye-ball detection of features characteristics of interactions, using the 
photometric SDSS-DR7. 
We defined two categories: $Disturbed$ and $Non\ disturbed$ pairs.
$Disturbed$ pairs are sub-classified as merging ($M$, pairs with evidence of an ongoing merging process)
and tidal ($T$, pairs with signs of tidal interactions but not necessarily merging).
$Non\ disturbed$ ($N$) pairs showing no evidence of distorted morphologies.
Fig. \ref{ej} shows images of typical examples of pair galaxies for  
different visual classification: $M$, $T$ and $N$. 
For the pair catalog we calculate the percentage of these three categories defined above.
As we can notice from Table 2, we find that about 10 $\%$ of galaxy pairs are classified as $M$, 
30 $\%$ as $T$ and 60 $\%$ as $N$, these percentages do not depend on redshift.

The visual inspection was performed by one of the authors in order to maintain a
 unified criteria. The reliability of the classification was addressed by the comparison with the classification of a subsample of pairs by another author. This procedure allows us to quantify the uncertainty in the classification (see table 2).

\begin{figure}
\includegraphics[width=70mm,height=100mm ]{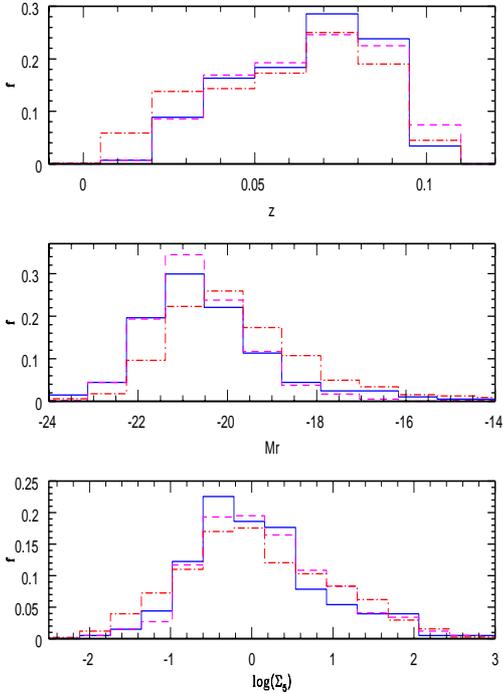}
\caption{Distribution of z, $M_r$ and log($\Sigma_5$) (upper, medium and lower panel, respectively) for pair galaxies classified as $M$, $T$ and $N$(solid, dashed and dot-dashed lines, respectively)  }
\label{histz1}
\end{figure}

We show in Fig.~\ref{histz1} the z, $M_r$ and log($\Sigma_5$)\footnote{$\Sigma_5$
is calculated through the projected distance, $d$, to the $5^{th}$ brightest nearest 
neighbor ($M_r < -20.5$) with a radial velocity difference lesser than 1000 
km $s^{-1}$, $\Sigma = 5/(\pi d^2)$.} distributions (upper, medium and lower panels, respectively) for $M$,
$T$ and $N$ systems (solid, dashed and dot-dashed lines, respectively). 
It can be appreciated that the galaxies of the different pair interaction categories show similar redshift 
trends.
The middle panel show that the luminosity distributions of the three interaction categories are comparable albeit with a small trend for higher luminosities in the $M$ and $T$ samples.
The lower panel shows that the  log($\Sigma_5$) distributions of  
$M$, $T$ and $N$ pairs are remarkably similar. This provides an important result regarding the local environment of galaxies undergoing merger process in the present universe.

\begin{table}
\center
\caption{Percentages of pairs classified as $M$, $T$ and $N$}
\begin{tabular}{|c c c| }
\hline
Classification & Number of pairs &  Percentages  \\
\hline
\hline
Merging              &  205      &    10.43$\%$ $\pm$ 0.5$\%$  \\
Tidal                &  589      &    30.03$\%$ $\pm$ 2.4$\%$    \\
Non Disturbed        &  1165     &    59.45$\%$ $\pm$ 4.2$\%$    \\
\hline
\end{tabular}
{\small}
\end{table}

\subsubsection{Dependence on $r_p$ and $\Delta V$}

\begin{figure}
\includegraphics[width=100mm,height=130mm ]{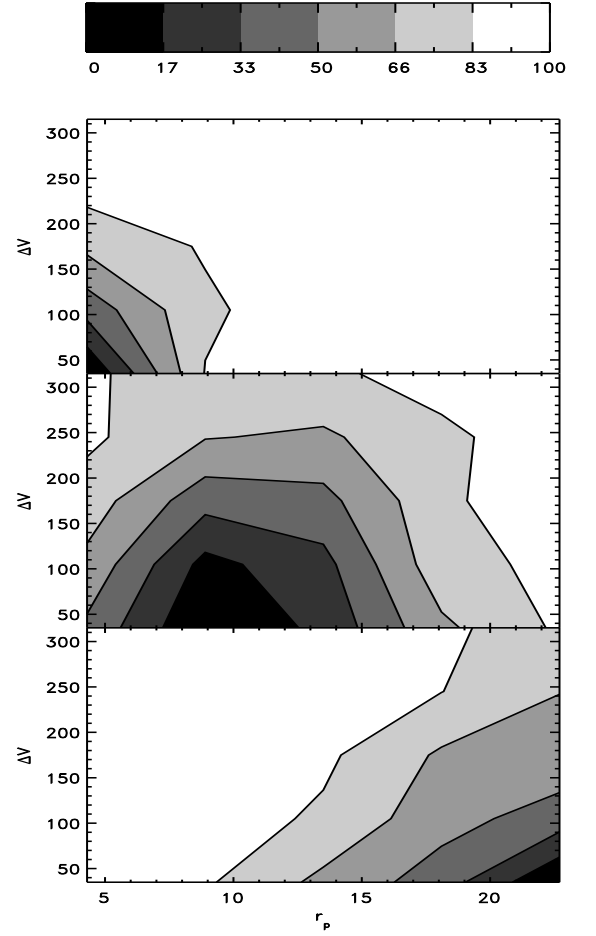}
\caption{Distribution of projected separation, $r_{\rm p}$, and relative radial velocity, $\Delta V$,
for $M$, $T$ and $N$ pairs (upper, medium and lower panels, respectively). 
The gray scale show the different percentages of enclosed pairs in a given contour.
}
\label{rpVMTN}
\end{figure}

In this subsection we analyze the dependence of pair classification on projected
distances, $r_p$, and relative velocities, $\Delta V$, between members.
For this purpose, we show in Fig. \ref{rpVMTN}  density contours in the $r_p$-$\Delta V$ plane for galaxies of 
the different interaction classes, $M$, $T$ and $N$.
The gray scale correspond to different 
percentages of pairs enclosed in a given contour.\\

As expected, given that the classification is based on visual appearance in projection,
 there is a trend for lower $r_p$ values for $M$ and $T$ types. Nevertheless, the classification cannot be reduced to a relative distance criterion and therefore a visual inspection of images is required to detect the interaction-driven morphological disturbances.
We find that the distribution of relative radial velocities is significantly lower in $M$ types as compared to $T$ and $N$ types.

\subsubsection{Comparison with the Galaxy Zoo Catalog}

We cross-correlate our sample of galaxy pairs with the galaxy zoo catalog (Lintott et al. 2011) to compare the two classification schemes. Galaxy Zoo comprises a morphological classification of nearly 900,000 galaxies drawn from the Sloan Digital Sky Survey, 
contributed by hundreds of thousands of volunteers in order
to cover a wide coverage  of the galaxy survey, however due to the large number of classifiers
it becomes complex to maintain a unified criteria and a reliable classification.
They define six categories (elliptical, spiral, spiral clockwise, spiral anticlockwise, merger or uncertain) and give the fraction of votes in each of the six categories. 
Objects classified as mergers are identified as galaxies with signs of collision.
We find 1417 common pairs in the two catalogs, where 596 pairs are classified as disturbed ($M$ or $T$), while only 128 objects were classified  as ``merger'' by the galaxy zoo team (fraction of votes $>0.5$). 

\begin{figure}
\includegraphics[width=85mm,height=85mm ]{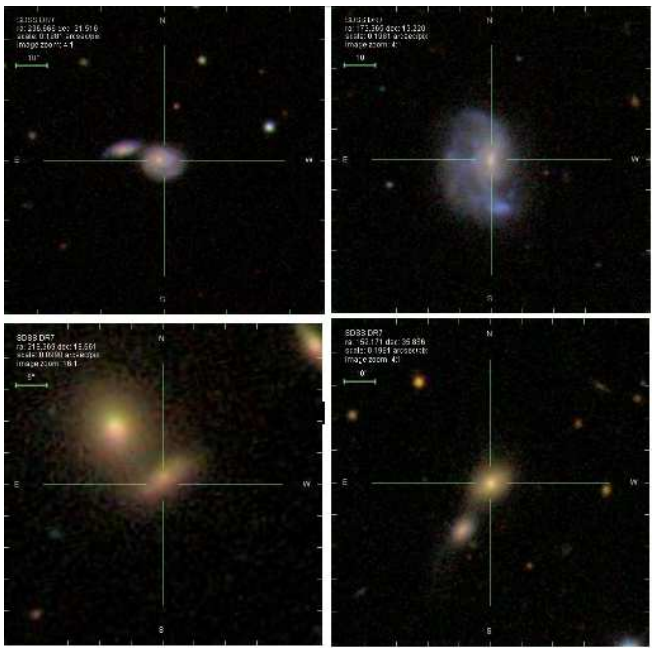}
\caption{Examples of pairs classified as $M$ or $T$.  For these galaxies, galaxy zoo provides a low fraction of 
votes for ``merger''.
The Figure shows the scale, size in arcsec and the N-E direction.
}
\label{zoo}
\end{figure}

 We show in Fig.~\ref{zoo} some typical examples of pairs we classified as $M$ and $T$, while galaxy zoo assigned an extremely low fraction of votes for a ``merger'' in these objects.

\section{Major and minor interactions}

It is expected that the effects of an interacting companion on a given object will strongly depend on their 
relative luminosity (mass proxy) ratio. For this reason, in this section we explore the dependence on the
luminosity ratio of the interaction-induced star formation activity and colors. This analysis may help to deepen our understanding of this issue which has been explored by different authors under diverse approaches.

Observational evidence (e.g. Donzelli \& Pastoriza 1997) shows that the faint members of an
interacting pair are more strongly affected by the companion. Nevertheless, in a previous 
work (Lambas et al. 2003), using a detailed statistical analysis on 2dF Galaxy Redshift Survey 
(2dFGRS, Colles et al. 2001) data, 
showed that the brightest component of a pair has the most enhanced star formation activity when compared 
to isolated galaxies of similar luminosity, suggesting that interactions may effectively trigger 
star formation on the brighter member pairs.
Ellison et al. (2008) found an enhancement of the SFR of galaxy pairs at projected separations $< 30-40 \rm \,kpc \,h^{-1}$, an effect that is stronger in major mergers.
More recently, Ellison et. al (2010) also found that both, the median mass ratio of pairs and the fraction of major-to-minor pairs, are independent of local environment.
 
In a similar way, Alonso et al. (2010), showed that galaxies with 
high stellar mass, low metallicity content and disturbed morphologies (characteristics of merger remnants) 
have bluer colors and younger stellar populations. These results would indicate that a close minor 
companion can induce significant inflows of external gas onto the central region which would lower the metallicity and trigger star formation in the most massive, morphologically disturbed galaxies.

\begin{figure}
\includegraphics[width=90mm,height=90mm ]{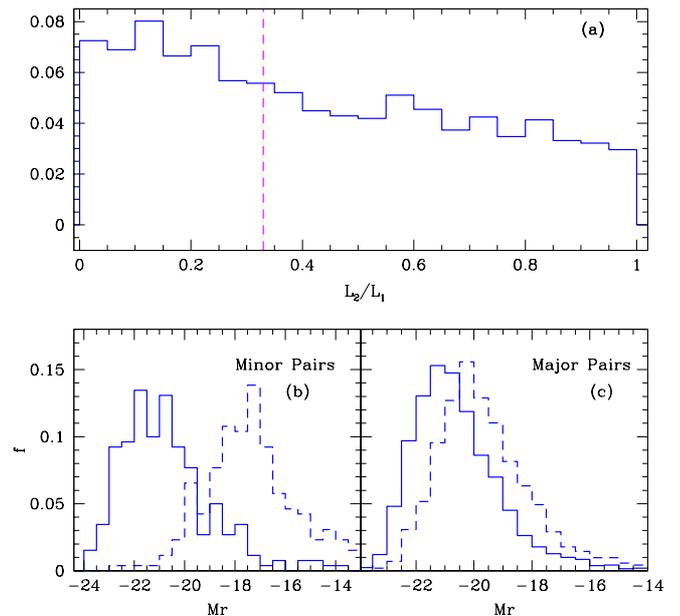}
\caption{$(a)$ Distribution of the luminosity ratios of galaxies in the pair sample.
$(b)$ and $(c)$ show the $M_r$ distributions of the most luminous (solid lines) and 
less luminous (dashed lines) galaxy member in minor and major pair subsamples, respectively.}
\label{HLum}
\end{figure}

For the present analysis we have divided our sample in major and minor 
interaction pairs according to the luminosity ratio of the galaxy members, the usually adopted criterium for 
the classification into major or minor interaction. 
In Fig.~\ref{HLum} ($a$) we show the distribution of the $L_2/L_1$ ratio, 
and the adopted threshold $L_2/L_1 =0.33$ which gives 877 minor and 1082 major interactions. 

\begin{figure}
\includegraphics[width=90mm,height=45mm ]{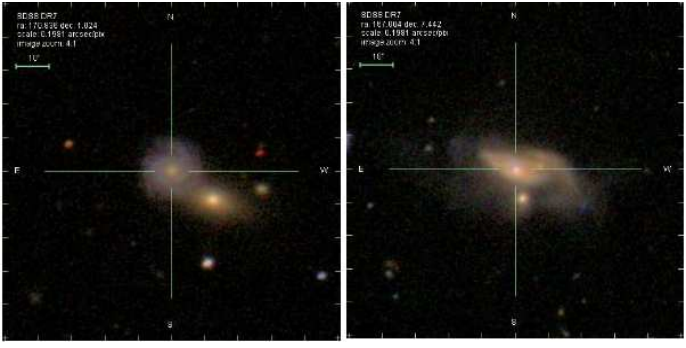}
\caption{Examples of major and minor interactions. Major interaction (left) with z=0.075 and $L_2/L_1$=0.14; minor interaction (right) with z=0.041 and $L_2/L_1$=0.51. The Figure shows the scale, size in arcsec and the N-E direction. }
\label{mm}
\end{figure}

The luminosity distributions of the galaxy members of these subsamples of pairs are shown in Fig.~\ref{HLum} ($b$ and $c$) and in Fig.~\ref{mm} we show some examples of major and minor galaxy encounters.
Table 3 shows the percentages of pairs classified as $M$, $T$ and $N$ in minor  
and major interactions where it can be seen their similarity regardless the relative luminosity ratio.

\begin{figure}
\includegraphics[width=90mm,height=130mm ]{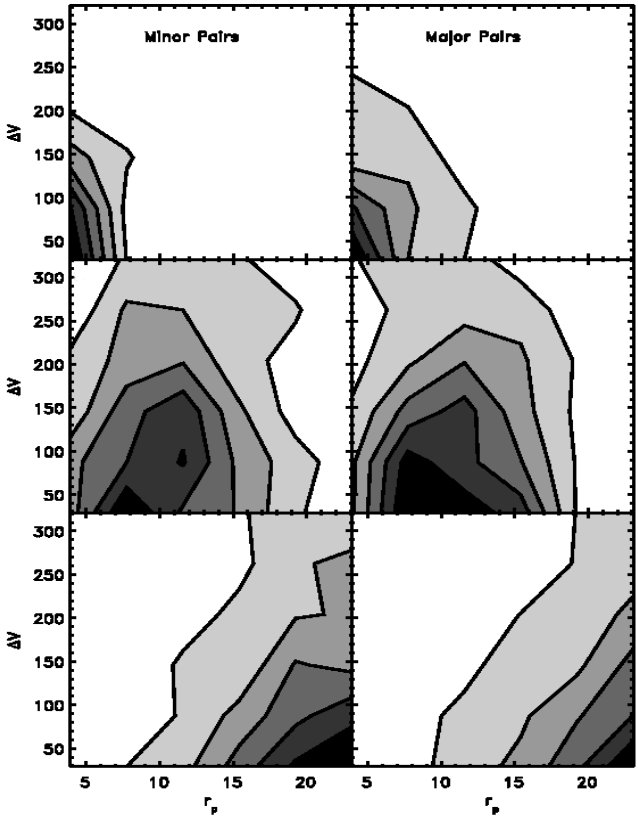}
\caption{Distribution of projected separation, $r_{\rm p}$, and relative radial velocity, $\Delta V$,
for $M$, $T$ and $N$ pairs (upper, medium and lower panels, respectively), in minor (left) and major (right) pairs. 
The grey scale show the different percentages of enclosed pairs in a given contour
(17$\%$, 33$\%$, 50$\%$, 66$\%$ and 83$\%$).
}
\label{rpVMTN1}
\end{figure}

Following section 2.1.1, we performed the same analysis of the dependence of pair classification on projected
distances, $r_p$, and relative velocities, $\Delta V$,  for the subsamples of major and minor pairs.
We show in Fig. \ref{rpVMTN1}  density contours in the $r_p$-$\Delta V$ plane for galaxies of 
the different interaction classes, $M$, $T$ and $N$, in major and minor interactions 
(left and right panels, respectively).
The gray scale correspond to different percentages of pairs enclosed in a given contour. 
It can be observed in this figure similar results as in Fig. \ref{rpVMTN}, indicating that  
the trends do not depend on the luminosity ratio of the galaxies.

\begin{table}
\center
\caption{Percentages of pairs classified as $M$, $T$ and $N$, in minor  
and major interaction pairs }
\begin{tabular}{|c c c| }
\hline
Classification & Number of minor pairs &  Percentages   \\
\hline
\hline
Minor Pairs          & 877  &     100$\%$        \\
\hline
Merging              &  102   &    11.63$\%$     \\
Tidal                &  260  &    29.64$\%$     \\
Non Disturbed        &  515   &    58.73$\%$     \\
\hline
\hline
Classification & Number of major pairs  &  Percentages  \\
\hline
\hline
Major Pairs          & 1082  &     100$\%$        \\
\hline
Merging              &  103  &    9.51$\%$     \\
Tidal                &  330  &    30.51$\%$    \\
Non Disturbed        &  649  &    59.98$\%$      \\
\hline
\end{tabular}
{\small}
\end{table}


\subsection{Galaxy colors}

\begin{figure}
\includegraphics[width=90mm,height=110mm ]{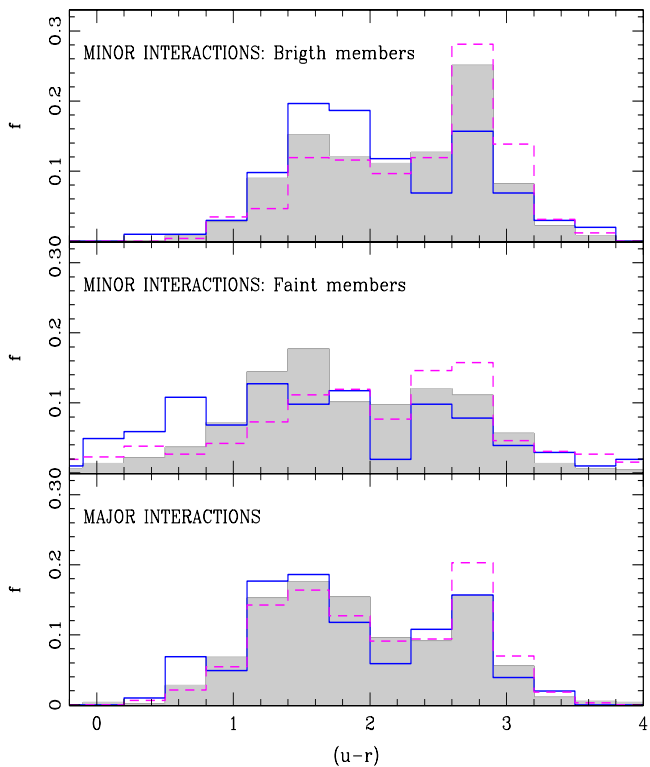}
\caption{Distribution of $u-r$ colors for the brightest and faintest pair members in minor interactions (upper and medium panels, respectively). Solid lines, dashed lines and full surface correspond to $M$, $T$ and $N$ pairs respectively. The lower panel correspond to major interactions.}
\label{ColMM}
\end{figure}

In order to explore the effects of galaxy interactions on the color index of the pair members,
in Fig.~\ref{ColMM} we show the $(u-r)$ color distributions of galaxies in major and minor interactions, indicating separately the bright and faint components in the minor interactions 
(upper and medium panels respectively).
Since the two galaxy members in major pairs have similar luminosity  
(see Fig.~\ref{HLum}), the division in this case is not important.  
It can be seen that there is an excess of galaxies in the blue peak in $M$ systems. 
In particular, the faint members of minor interaction pair show a significant fraction of galaxies with extremely blue colors ($u-r <$ 1). 
Similar results were obtained by Woods \& Geller (2007), who found that the faint members in minor pairs show enhanced star formation activity. 

On the other hand, $T$ pairs show an excess of galaxies in the red peak
as compared to $N$-types.

The absence of an intermediate color population indicates that the process responsible for the transformation from blue to red colors needs to be very fast and efficient (e.g. Baldry et al. 2004, Balogh et al. 2004).
In this context, our results suggest that the variation of the blue and red peak locations of the color bimodal distribution could be driven by different aspects of galaxy interactions such as evolutionary stage, gas content, interaction strength, etc.

\subsection{Galaxy 4000 $\AA$ discontinuity}

In the following analysis we use the spectral index $D_n(4000)$, 
as an indicator of the age of stellar populations. 
This spectral discontinuity occurring at 4000 $\AA $ (Kauffmann et al. 2003) arises by an
accumulation of a large number of spectral lines in a narrow region of the spectrum, an effect that is important in the spectra of old stars.
We have adopted Balogh et al. (1999) definition of  $D_n(4000)$  as the ratio of the average flux densities in the narrow continuum bands (3850-3950 $\r{A}$ and 4000-4100 $\r{A}$). 

\begin{figure}
\includegraphics[width=90mm,height=110mm ]{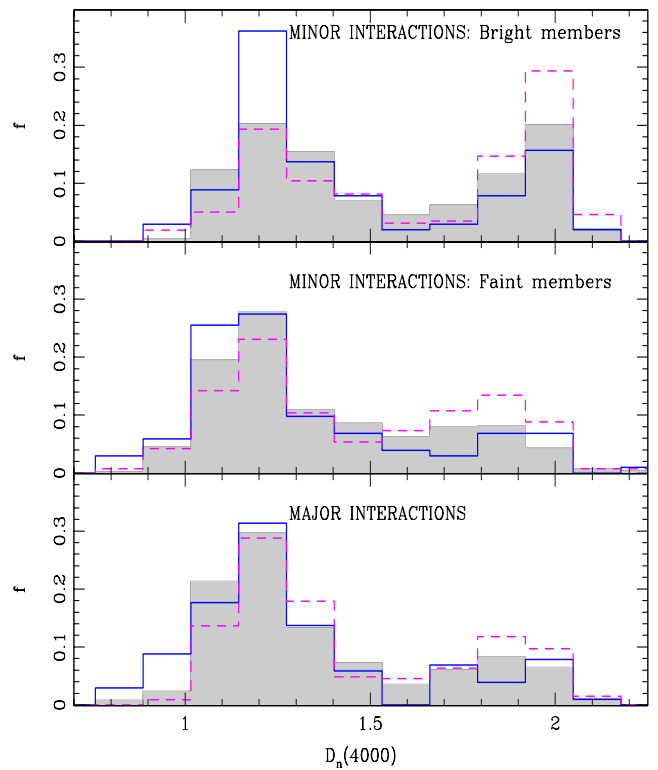}
\caption{Distribution of $D_n(4000)$ for the brightest and faintest galaxy members in minor pairs, upper and medium panels respectively. Solid and dashed lines correspond to $M$ and $T$ types respectively, full surface correspond to $N$ types.
In the lower panel we show the corresponding distributions for major interactions.  }
\label{DnMM0}
\end{figure}

In Fig.~\ref{DnMM0} we show the distribution of $D_n(4000)$ values for galaxies in $M$, $T$ and $N$ types.
We also separate in this figure the results for the brightest and the faintest members in minor interactions. 
In agreement with the previous results for colors, we find an excess of $M$ types exhibiting  low  values of D$_n$(4000) showing that galaxies undergoing strong 
interactions are dominated by young stellar populations, in agreement with Woods et al. (2010).

We also find evidence that $T$ pairs have an excess of large $D_n(4000)$ values, characteristic of
old stellar populations.

Possible explanations for  the fact that pairs with tidal features have a higher fraction of old stellar populations (reflected by both $u-r$ and $D_n(4000)$ distributions) can be related to the longer timescale of morphological disturbance with respect to that of the tidally induced star formation. Thus, these pairs may have an aged (reddened) stellar population and still present strong signs of a past interaction. Besides,
it can be argued that strong tidal features can be associated to disruptive effects present in some tidal interactions of galaxy disks which would lead to lower gas densities and therefore lower star formation rates in these systems.

\begin{figure}
\includegraphics[width=70mm,height=100mm ]{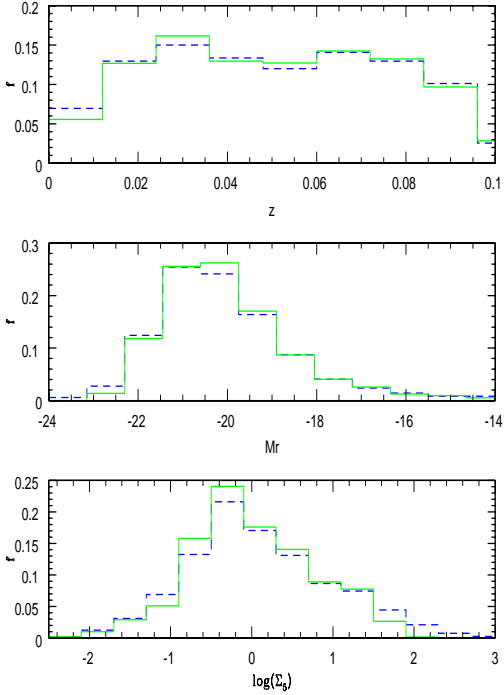}
\caption{Distribution of z, $M_r$ and log($\Sigma_5$) in pair galaxies (dashed lines) and in the control sample (solid lines)  }
\label{histz}
\end{figure}

In order to provide a suitable quantification of the effects of strong interactions on the relative fraction of star-forming galaxies at a given luminosity we have considered a single threshold in $D_n(4000)=1.5$ to divide
 the sample into star-forming and passive galaxies. For comparison, we also construct  a control sample for the pair catalog,
defined by galaxies without a close companion within the adopted separation 
and velocity thresholds.  
By using a Monte Carlo algorithm, for each galaxy pair, we selected 
two other galaxies without a companion within $r_{p}< 100 \rm \,kpc \,h^{-1}$
and relative radial velocities, $\Delta V < 350 \rm \,km \,s^{-1}$. 
Moreover, these galaxies were also required to match the observed 
redshift, luminosity and local density environment, $\Sigma_5$,
distributions of the corresponding pair sample, to represent 
a robust control sample (Perez et al. 2009).
In Fig.\ref{histz} we show the redshift, $M_r$ and log($\Sigma_5$) distributions
(upper, medium and lower panels, respectively) for pair galaxy catalog (dashed lines) 
and their corresponding control sample (solid lines). 
We have also explored these distributions with the restriction $D_n(4000)<1.5$ in
both pair and control samples, finding similar distributions of redshift, luminosity, and local density environment. 
Taking into account this result, we conclude that this set of control galaxies is suitable for our purpose. Furethermore,
we have also built a second test sample of close galaxies in projection ($r_{p} < 30 \rm \,kpc \,h^{-1}$), but with a large redshift difference ($\Delta V > 2000 \rm \,km \,s^{-1}$). We find that this sample of unphysical pairs behaves remarkably similar to the control sample in the different analysis performed. This gives another indication for the reliability of the control sample and that the spatial proximity is a crucial parameter that determines the behaviour of pair galaxies.

\begin{figure}
\includegraphics[width=90mm,height=110mm ]{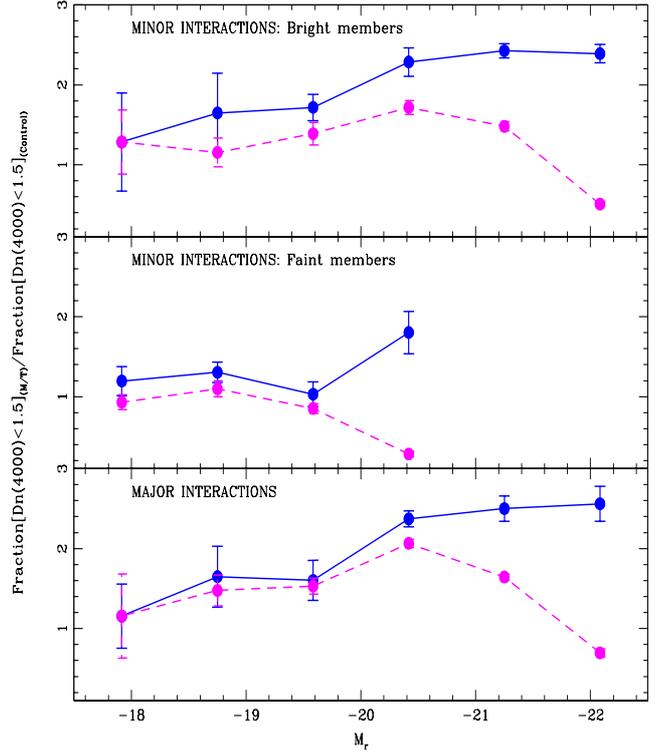}
\caption{The fraction of galaxies with $D_n(4000)<1.5$ relative to the control sample,
fraction ($D_n(4000)<1.5_{(M/T)}$)/fraction ($D_n(4000)<1.5_{(Control)}$, as a function of 
$M_r$, for the brightest and the faintest pair members in minor interactions, upper and medium panels respectively. Solid and dashed lines correspond to $M$ and $T$ types, respectively.
In the lower panel we show the corresponding distributions for major interactions. 
The errors shown were calculated within uncertainties derived through the bootstrap re-sampling technique.
 }
\label{DnMM}
\end{figure}

In Fig.~\ref{DnMM} we show the fraction of star-forming galaxies, $D_n(4000)< 1.5$, as a function of $M_r$ in $M$ and $T$ types relative to the control sample (fraction ($D_n(4000)<1.5_{(M/T)}$)/fraction ($D_n(4000)<1.5_{(Control)}$).
In this figure we distinguish the brightest and faintest members in minor interactions (upper and medium panels, respectively), the lower panel correspond to major interactions. 
We find that for the more luminous systems the effects of interactions in $M$ pairs are strongest (by a factor $\approx$ 2) in terms of the relative fraction of star-forming galaxies. 
In close galaxy pairs Ellison et al. (2008, 2010) and 
Woods \& Geller (2007) show similar trends. 
However, in $T$ pairs, there is an opposite trend, with lower amplitude.

\begin{figure}
\includegraphics[width=90mm,height=110mm ]{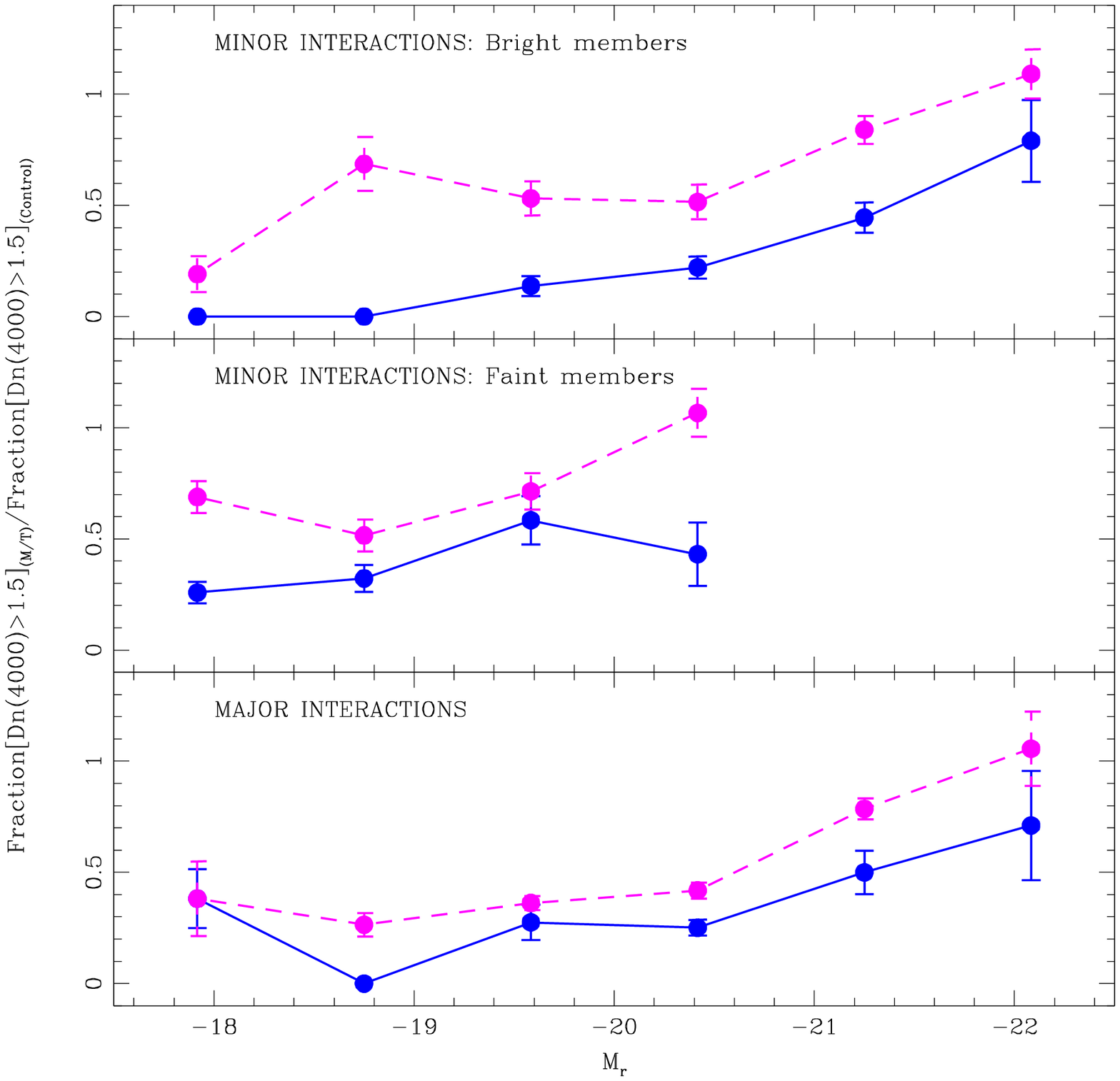}
\caption{The fraction of galaxies with $D_n(4000)>1.5$ relative to the control sample,
fraction ($D_n(4000)>1.5_{(M/T)}$)/fraction ($D_n(4000)>1.5_{(Control)}$, as a function of 
$M_r$, for the brightest and the faintest pair members in minor interactions, upper and medium panels respectively. Solid and dashed lines correspond to $M$ and $T$ types, respectively.
In the lower panel we show the corresponding distributions for major interactions. 
The errors shown were calculated within uncertainties derived through the bootstrap re-sampling technique. 
 }
\label{DnoldMM}
\end{figure}

We have also explored the dependence of the fraction of the old stellar population in galaxies of  $M$ and $T$ types, (solid and dashed lines, respectively), relative to the control sample 
(fraction ($D_n(4000)>1.5_{(M/T)}$)/fraction ($D_n(4000)>1.5_{(Control)}$) as a function of $M_r$. 
 Fig.~\ref{DnoldMM} shows an equivalent set of plots to those of the young stellar population fractions displayed 
in Fig.~\ref{DnMM}, also considering separately major interactions, and the brightest and faintest members in minor interactions. 
It can be seen in Fig.~\ref{DnoldMM} an increase of the old stellar population fraction in luminous systems 
and that the effects of interactions in $T$ pairs are strongest. This provides further support to our hipothesis 
of disruptive effects of some interactions that could lead to both, our classification into the $T$ class, and an associated larger fraction of old stellar populations provided by a shutdown of the provision of gas to form new stars.

\subsection{Global star formation efficiency in major and minor interactions}

Since interacting galaxies may finally end in a single system, in this subsection we analyze 
the efficiency of interactions to trigger the formation of stars in the pair considered as a whole.

\begin{figure}
\includegraphics[width=80mm,height=100mm ]{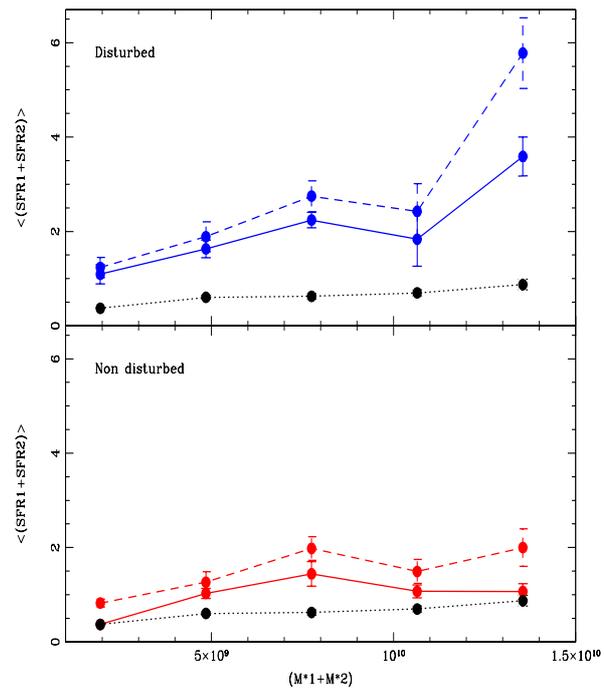}
\caption{Total star formation rate $<SFR_1+SFR_2>$ as a function of total stellar mass 
$M^*_1+M^*_2$ for major (dashed) and minor (solid) interactions classified as Disturbed 
($M$ and $T$) and non-disturbed ($N$) (upper and lower panels, respectively). 
Dotted lines represent the control sample within 
uncertainties derived through the bootstrap re-sampling technique. }
\label{SFR12M12}
\end{figure}

For this aim, we compute the sum of the stellar masses and the sum of the star formation 
rates for the two members of a given pair using the data given in Brinchman et al (2004).
Fig.~\ref{SFR12M12} shows the behavior of the total star formation rate ($SFR_1 + SFR_2$) as 
a function of the total stellar mass ($M^*_1 +M^*_2$).
It is clearly seen by comparison of the upper and lower panels that in both, minor and major 
interactions, pairs with tidal signatures (M and T pairs) have a significantly higher total star 
formation rate than N pairs. The comparison with a control sample shows that  
interactions show enhanced star formation activities.
It can also be appreciated that, at a given total stellar mass and irrespective of the 
morphological appearance of the interacting pairs, major interactions are those more 
efficient in forming new stars (up to a factor 2).

\begin{figure}
\includegraphics[width=80mm,height=100mm ]{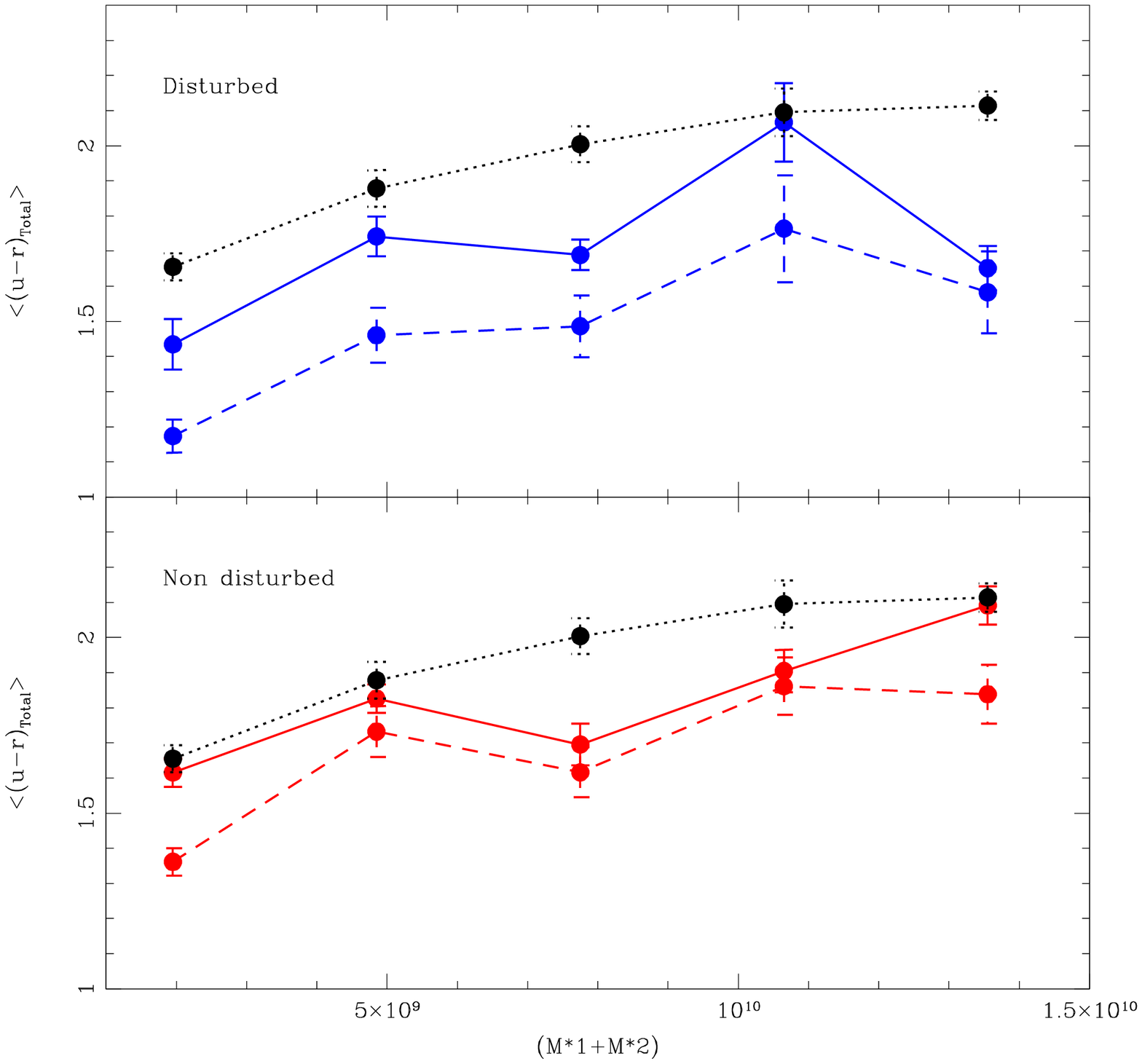}
\caption{Total $<(u-r)>$ as a function of total stellar mass 
$M^*_1+M^*_2$ for major (dashed) and minor (solid) interactions classified as Disturbed 
($M$ and $T$) and non-disturbed ($N$) (upper and lower panels, respectively). 
Dotted lines represents the control sample within 
uncertainties derived through the bootstrap re-sampling technique. }
\label{Col12M12}
\end{figure}

In a similar way, we performed an analysis computing the global index colors 
as a function of a total stellar mass. The results are displayed in
Fig.~\ref{Col12M12}.  
It can be seen that galaxies in disturbed pairs (upper panel) show bluer colors than
non-disturbed systems (lower panel), at a given total stellar mass;
and we have also noted that major mergers show a significant bluer population. 
Also, galaxies in the control sample are redder than galaxies in different interaction 
classes (minor/major mergers, disturbed/non-disturbed pairs),  
indicating that the global efficiency of different stages/classes of  
interactions are associated with triggered star formation activity, reflected in the blue colors.

\section{Summary and Conclusions}

We have performed a statistical analysis of 1959 galaxy pairs ($r_p < 25$ kpc $h^{-1}$ and 
$\Delta V < 350$ km $s^{-1}$) within $z<0.1$ selected from SDSS-DR7 and we
have carried out an eye-ball classification of images according to the evidence of
interaction through distorted morphologies and tidal features.

We can summarize the main results in the following conclusions.

\begin{itemize}

\item   We classified $10\%$ of the total pair sample as merging,  
$30\%$ with tidal features, and $60\%$ as non disturbed. 
We also explore the relation between projected separation and relative radial velocity, 
showing that $r_p$ and $\Delta V$ ranges overlap for the different pair categories.
This result indicates that neither relative distance nor radial velocity between pair members are
enough to predict the interaction class assignment, so that visual inspection of images is required
 to properly classify galaxy interactions.

\item We separate the pair galaxy catalog into minor and major mergers according to the relative 
luminosities of the galaxy members and consider galaxy colors. We find significant changes in the color distribution according to 
the relative luminosity of the pair members.
We find that the bright and faint members in $M$ minor interactions are bluer than those in $T$ and $N$ pairs. 
We notice that this tendency is more important in the faintest galaxy pair members.
$T$ systems show a large population of red galaxies with respect to $N$-types. 

Therefore, these results suggest that galaxy interactions are important in driving 
the evolution of galaxy color bimodality mainly through an induced inflow of gas  
forming new star generations and the effects of tidal disruption.

\item We have considered a single threshold in $D_n(4000)=1.5$ to divide
 the sample into star-forming and passive galaxies. The $D_n(4000)$ distributions 
also show an excess of young stellar population in $M$ pairs, indicating recently triggered star 
formation events. In agreement with colors, we find that $T$ pairs show a significant excess of
old stellar populations.
We find that the relative fraction of luminous star-forming galaxies in $M$ pairs is higher by a factor $\approx$ 2 as compared to the control sample.

\item We have also performed an analysis of the pairs considered as a single system. We find that
at a given total stellar mass, major interactions are more efficient in forming new stars
in comparison to minor pairs (by a factor $\approx$ 2). Nevertheless, in both, minor 
and major interactions, disturbed pairs ($M$ and $T$ systems) have a significantly 
higher total star formation rate than non-disturbed galaxies. 
In a similar way, at a given total stellar mass, disturbed pairs show 
blue global color with respect to non-disturbed systems.

\end{itemize}

We conclude that galaxy interactions and mergers provide key mechanisms that regulate galaxy properties.
We find that encounters of galaxies with similar luminosities are globally more effective in forming new stars, in comparison to minor mergers. We also show that this process is significantly more efficient in pairs with strong signs of interactions.
Finally, we argue that the ratio between the luminosity of the galaxy members involved in a merger, and the characteristics of interactions are important issues in setting star formation activity, building the stellar populations and global galaxy colors.

\begin{acknowledgements}
      This work was partially supported by the Consejo Nacional de Investigaciones
Cient\'{\i}ficas y T\'ecnicas and the Secretar\'{\i}a de Ciencia y T\'ecnica 
de la Universidad Nacional de San Juan.

Funding for the SDSS has been provided by the Alfred P. Sloan
Foundation, the Participating Institutions, the National Science Foundation,
the U.S. Department of Energy, the National Aeronautics and Space
Administration, the Japanese Monbukagakusho, the Max Planck Society, and the
Higher Education Funding Council for England. The SDSS Web Site is
http://www.sdss.org/.

The SDSS is managed by the Astrophysical Research Consortium for the
Participating Institutions. The Participating Institutions are the American
Museum of Natural History, Astrophysical Institute Potsdam, University of
Basel, University of Cambridge, Case Western Reserve University,
University of
Chicago, Drexel University, Fermilab, the Institute for Advanced Study, the
Japan Participation Group, Johns Hopkins University, the Joint Institute for
Nuclear Astrophysics, the Kavli Institute for Particle Astrophysics and
Cosmology, the Korean Scientist Group, the Chinese Academy of Sciences
(LAMOST), Los Alamos National Laboratory, the Max-Planck-Institute for
Astronomy (MPIA), the Max-Planck-Institute for Astrophysics (MPA), New Mexico
State University, Ohio State University, University of Pittsburgh, University
of Portsmouth, Princeton University, the United States Naval Observatory, and
the University of Washington.
The authors thank Dr. Nelson Padilla for a detailed revision and useful comments.
\end{acknowledgements}

\end{document}